\documentclass[aps,12pt,eqsecnum]{revtex4}
\usepackage{amsmath}
\usepackage{graphicx}
\newcommand{\beq}{\begin{equation}}
\newcommand{\eeq}{\end{equation}}
\newcommand{\bes}{\begin{subequations}}
\newcommand{\ees}{\end{subequations}}
\newcommand{\bea}{\begin{eqnarray}}
\newcommand{\eea}{\end{eqnarray}}

\newcommand{\ro}{\mbox{\boldmath$\rho$}}
\newcommand{\robs}{(\ro-{\bf b})^2}
\newcommand{\zeze}{{\hat {\bf z}}}
\newcommand{\piro}{{_{\rm P}}}
\newcommand{\pici}{{_{\rm C}}}
\newcommand{\piel}{{_{\rm L}}}
\newcommand{\piti}{{_{\rm T}}}

\begin{document}
\title{Gauge dependence of calculations in relativistic Coulomb 
excitation}
\author{B.F. Bayman\\
{\it School of Physics and Astronomy, University of Minnesota,}\\
{\it 116 Church Street S.E., Minneapolis, MN 55455, U.S.A.}\\
and\\
F. Zardi\\
{\it Istituto Nazionale di Fisica Nucleare and Dipartimento di Fisica, \\}
{\it Via Marzolo,8 I-35131 Padova,Italy.}}

\begin{abstract}
Before a quantum-mechanical calculation involving electromagnetic interactions 
is performed, a choice must be made of the gauge to be used in expressing the 
potentials. If the calculation is done exactly, the observable results it 
predicts will be independent of the choice of gauge. However, in most practical
 calculations approximations are made, which can destroy the gauge invariance 
of the predictions. We compare here the results of coupled-channel 
time-dependent relativistic Coulomb excitation calculations, as performed in 
either Lorentz or Coulomb gauges. We find significant differences when the 
bombarding energy per nucleon is $\geq$ 2 GeV, which indicates that the common 
practice of relying completely on the Lorentz gauge can be dangerous. 
\end{abstract}

\maketitle
\date{\today}

\section{Introduction}
Coulomb excitation is a collison process in which the predominant 
projectile-target interaction is electromagnetic. The electromagnetic 
interaction is expressed in the Hamiltonian, and therefore also in the 
Schr\"odinger equation, in terms of the electromagnetic potentials, 
$\left(\varphi({\bf r},t), {\bf A(r},t)\right)$. The potentials are subject 
to gauge ambiguity, since they are not uniquely determined by the charge and 
current distributions that create the electromagnetic field. Equivalently, 
if a gauge transformation generated by $\Lambda({\bf r},t)$ is performed, 
the new potentials $\left(\varphi'({\bf r},t), {\bf A'(r},t)\right)$ defined by
\bes
\bea
\varphi'({\bf r},t) ~&\equiv&~\varphi({\bf r},t)-\frac{1}{c}\frac{\partial 
\Lambda({\bf r},t)}{\partial t}\\
\bf A'({\bf r},t)~&\equiv& ~{\bf A(r},t)+\nabla \Lambda({\bf r},t)
\eea
\ees
yield the same 
$\left({\bf E(r},t), {\bf B(r},t)\right)$ as did $\left(\varphi({\bf r},t), 
{\bf A(r},t)\right)$ \cite{JAC}. Thus $\left(\varphi({\bf r},t), {\bf A(r},t)
\right)$ and $\left(\varphi'({\bf r},t), {\bf A'(r},t)\right)$ are both 
consistent with Maxwell's equations and the given charge and current 
distributions. Since the Schr\"odinger equation depends explicitly on the 
potentials, it is changed by a gauge transformation, and so is the wave 
function which is the solution of the Schr\"odinger equation. However, when 
an observable quantity (such as a transition probability) is calculated from 
this wave function, the result is invariant under a gauge transformation, 
even though the wave function is not.

The gauge invariance of an observable applies only if the exact solution of 
the Schr\"odinger equation is used in its calculation. If an approximate 
solution is used, then the calculated observable quantity may or may not be 
gauge invariant. A non-gauge invariant prediction is a serious drawback for 
a theory, since there is generally no \textit{a priori} reason to choose one 
gauge rather than another.

An example of an approximate solution that \textit{is} gauge invariant is 
given by first-order time-dependent perturbation theory\cite{WA}. Suppose 
we want to estimate transition probabilities in the target under the 
influence of a time-dependent external potential $V({\bf r},t)$ provided by 
the  projectile. If we use first-order time-dependent perturbation theory, 
then the predicted transition probability from target state 
$\phi_\alpha$ at $t=-\infty$ to target state $\phi_\gamma$ at $t=\infty$ 
is given by
$$
\left|~\int_{-\infty}^\infty \frac{d t}{i \hbar}~
e^{i \omega_{\gamma \alpha}t}~V_{\gamma \alpha}(t)~\right|^2,
$$
where
\bes
\bea 
\omega_{\gamma \alpha}&\equiv&(\epsilon_\gamma~-~\epsilon_\alpha)/\hbar\\
V_{\gamma \alpha}(t)&\equiv& <\phi_\gamma~|~V({\bf r},t)~|~\phi_\alpha>.
\eea
\ees
Thus the first-order transition probability is proportional to the square of 
the modulus of the $\omega=\omega_{\gamma \alpha}$ (onshell) Fourier 
component of $V_{\gamma \alpha}(t)$. If $\rho_{\gamma \alpha}$ and 
${\bf J}_{\gamma \alpha}({\bf r},t)$ are the target transition charge and 
current densities associated with the $\phi_\alpha \rightarrow \phi_\gamma$ 
transition, the electromagnetic interaction $V_{\gamma \alpha}(t)$ is 
\beq
V_{\gamma \alpha}(t)=\int d^3r~\left[~\rho_{\gamma \alpha}
({\bf r})\varphi({\bf r},t)-\frac{1}{c}{\bf J}_{\gamma \alpha}({\bf r})
\cdot {\bf A(r},t)~\right].
\eeq
Target charge conservation can be expressed by the relation
\beq 
\nabla \cdot {\bf J}_{\gamma \alpha}({\bf r},t)~+~i 
\omega_{\gamma \alpha}\rho_{\gamma \alpha}({\bf r})~=0.
\eeq
This can be used to show that the $\omega$ Fourier component of 
$V_{\gamma \alpha}(t)$, 
\bes
\bea
V_{\gamma \alpha}(\omega)~&\equiv&~\int_{-\infty}^\infty 
\frac{d t}{i \hbar}~e^{i \omega t}~V_{\gamma \alpha}(t)\\
~&=&~\int_{-\infty}^\infty \frac{dt}{i \hbar}e^{i \omega t}
\int d^3r~\left[~\rho_{\gamma \alpha}({\bf r})\varphi({\bf r},t)-\frac{1}{c}
{\bf J}_{\gamma \alpha}({\bf r},t)\cdot {\bf A(r},t)~\right],
\eea
\ees
is changed by the gauge transformation\footnote{We restrict our attention to 
gauge generators that vanish at $t=\pm \infty$.}  
generated by $\Lambda({\bf r},t)$ by
$$
\frac{(\omega-\omega_{\gamma \alpha})}{\hbar c}\int_{-\infty}^{\infty}dt 
\int d^3r e^{i \omega t}\rho_{\gamma \alpha}({\bf r})\Lambda({\bf r},t).
$$
Thus the onshell, $\omega=\omega_{\gamma \alpha}$, Fourier component 
$V_{\gamma \alpha}(\omega_{\gamma \alpha})$, which determines the first-order 
transition probability, is unchanged by a gauge transformation.

There are other approximations which depend only on the onshell Fourier 
transform of the interaction. For example, in the model in which the target 
is represented by a harmonic oscillator and the interaction is assumed 
to be linear in the oscillating variable or its conjugate momentum, an exact 
solution of the time-dependent Schr\"odinger equation is available\cite{MER}. It depends 
only on the onshell Fourier transform of the time-dependence of the 
interaction. Also, the FWW ``method of virtual quanta" (see, e.g., 
Reference\cite{BerBau}) is essentially first-order perturbation theory, in 
which the perturbing electromagnetic field is approximated by a plane wave. 
In both theories, the gauge independence of the onshell Fourier transform of 
the interaction guarantees that their predictions are gauge invariant.  

However this gauge invariance is not guaranteed if the full electromagnetic 
interaction is used, and if one wishes to go beyond first-order perturbation 
theory. For example, a coupled-channel approach to the time-dependent 
Schr\"odinger equation has been used
in an attempt to get a better description of Coulomb excitation to 
multi-phonon states of giant resonances
\cite{BerBau,EML,ChFr,BCH,LCC,ABE,BerPon99,BZ}. One begins by choosing a set 
of target states which can be expected to play a significant role in the 
reaction, and then solving the Schr\"odinger equation within the Hilbert 
space defined by these states. This coupled-channel approach depends upon 
the entire $\omega$-dependence of $V_{\gamma \alpha}(\omega)$, not only 
on its value for $\omega=\omega_{\gamma \alpha}$. If all the target states 
were included in the calculation, then the coupled-channel solution of the 
Schr\"odinger equation would be exact, and calculated observables would be 
gauge invariant. However if, as is always the case, the set of target states 
is truncated to make the calculation feasible, then the result of the 
coupled-channel calculation is approximate, and therefore calculated 
observables are not gauge invariant. The main concern of this paper is 
with the lack of gauge invariance of the predictions of coupled-channel 
time-dependent solutions of the Schr\"odinger equation governing relativistic 
Coulomb excitation. Some related discussions of this subject have been given 
by Baltz, 
Rhoades-Brown, and Weneser \cite{BRW}, Rumrich and Greiner \cite{RG} and 
Kobe and Kennedy \cite{KK}.

The classic paper on relativistic Coulomb excitation \cite{WA} by A. Winther 
and K. Alder (WA) described the electromagnetic influence of the projectile 
on the target using classical electromagnetic fields in the Lorentz gauge. 
Since the main concern of WA was with first-order time-dependent perturbation 
theory, their calculated transition probabilities and cross-sections were 
gauge invariant. Subsequent workers in this field who used coupled-channel 
methods continued to use classical fields and the Lorentz gauge. However, 
as explained above, the extent to which the results of these calculations 
are gauge invariant is not certain. 

When studies are done in which the electromagnetic field is quantized, it is 
common to use the Coulomb (or transverse) gauge. The reason is that the field 
quanta (photons) will then be purely transverse. Table I compares the 
properties of the potentials in the Lorentz and Coulomb gauges.
\begin{table*}
\caption{\label{tab:table1}Comparison of the corresponding properties of 
the potentials in the Lorentz and Coulomb gauges, due to the projectile 
charge density, $\rho_\piro({\bf r},t)$ and current density, 
${\bf J}_\piro({\bf r},t)$.}
\begin{ruledtabular}
\begin{tabular}{cc}
Lorentz gauge&Coulomb gauge\\
$\nabla {\bf \cdot A}^\piel~+~\frac{1}{c}\frac{\partial}
{\partial t}\varphi^\piel~=~0$&$\nabla {\bf \cdot A}^\pici~=~ 0$\\
$\nabla^2 \varphi^\piel~-~\frac{1}{c^2}\frac{\partial^2}
{\partial t^2}\varphi^\piel~=~ -4 \pi \rho_\piro$&$\nabla^2 
\varphi^\pici~=~-4 \pi \rho_\piro$\\
$\nabla^2{\bf A}^\piel~-~\frac{1}{c^2}\frac{\partial^2 {\bf
A}^\piel}{\partial t^2}~=~-\frac{4 \pi}{c} {\bf J}_\piro$&$
\nabla^2{\bf A}^\pici-\frac{1}{c^2}\frac{\partial^2 {\bf
A}^\pici}{\partial t^2}-\frac{1}{c}\frac{\partial}
{\partial t}\nabla
\varphi^\pici~=-\frac{4 \pi}{c} {\bf J}_\piro$
\end{tabular}
\end{ruledtabular}
\end{table*}

We will consider two models for the radial density of the projectile, as seen 
in its own rest frame: a 
finite spherically-symmetric charge distribution, and a point charge. The 
expressions for the finite charge distribution will be presented as Fourier 
transforms, which are convenient for the multipole expansions needed when 
the initial and final nuclear states are angular momentum eigenstates. 
The point charge expressions are presented as functions of $t$, which can 
be easily interpreted. Numerical calculations show that there is very little 
difference between the predictions of the two models. 

In Section II we will find the generator $\Lambda({\bf r},t)$ 
which will take us from the commonly used Lorentz gauge to the Coulomb 
gauge in the classical-field approach to relativistic Coulomb excitation. 
This is already known for a point projectile\cite{BRW}, but we will derive 
the expression appropriate to a projectile of finite size. Section III
presents the interaction potentials calculated in these two gauges. 
In Section IV we compare multipole expansions of these potentials, and 
show that the Coulomb gauge potential is free of a divergence that appears 
in the Lorentz gauge potential at high bombarding energy. Sections V and VI 
apply these formulae to the excitation of multiphonon states of the giant 
dipole resonance in $^{40}$Ca, as a result of bombardment by 
$^{208}$Pb nuclei. Section VII presents our conclusions and some general observations about gauge invariance.

\section{The gauge transformation connecting the Lorentz and Coulomb gauges}

We follow the standard approach to relativisitic Coulomb excitation as 
proposed by WA. The projectile nucleus is assumed to travel along a 
straight-line orbit parallel to the ${\bf
{\hat z}}$ axis, with impact parameter ${\bf b}$, at constant speed $v$.
The magnitude of the impact parameter is large enough so that nuclear
interactions between the target and projectile are negligible.  Because
of the assumed large projectile momentum, the electromagnetic impulse
the projectile receives due to its interaction with the target has
little effect on its trajectory, so the projectile maintains its
constant speed and impact parameter throughout the collision. As the
projectile passes, the target nucleus feels the time-dependent
projectile electromagnetic fields, which induce transitions between the
quantum states of the target.

We seek the gauge function $\Lambda({\bf r},t)$ which generates the gauge 
transformation
\bes
\bea
{\bf A}^\pici({\bf r},t)&=&{\bf A}^\piel({\bf r},t)~+\nabla\Lambda({\bf r},t)
\\
\varphi^\pici({\bf r},t)&=&\varphi^\piel({\bf r},t)~-~\frac{1}{c}
\frac{\partial}{\partial t}\Lambda({\bf r},t)
\eea
\ees
between the potentials satisfying the Lorentz and Coulomb conditions listed 
in Table 1.

\subsection{Point projectile of charge $Z_{{\rm P}}e$.}
For a point projectile of charge $Z_\piro e$, we have a charge density given by
\beq
\rho_{{\rm P}}({\bf r},t)=Z_{{\rm P}}e~\delta({\bf r - b -{\hat z}}vt)=
Z_{{\rm P}}e~\delta\left(\ro-{\bf b +{\hat z}}(z-vt)\right).
\eeq
If this charge density is used on the right-hand sides of the equations for 
$\varphi$ in the second row of Table I, we get the the solutions
\bes
\bea
\varphi^\piel({\bf r},t)&=&\frac{Z_\piro e}{\sqrt{\frac{\robs}{\gamma^2}
+(z-vt)^2}}\\
\varphi^\pici({\bf r},t)&=&\frac{Z_\piro e}{\sqrt{\robs+(z-vt)^2}.}
\eea
\ees
Baltz, Rhoades-Brown and Weneser\cite{BRW} used these potentials and 
Equation (2.1b) to obtain
\bea
\Lambda({\bf r},t)&=&c\int_{-\infty}^t
\left[\varphi^\piel({\bf r},t')-\varphi^\pici({\bf r},t')\right]dt' \nonumber\\
&=&Z_\piro ec\int_{-\infty}^t\left[\frac{1}{\sqrt{\frac{\robs}{\gamma^2}
+(z-vt')^2}}~-~\frac{1}{\sqrt{\robs+(z-vt')^2}}\right]dt'\nonumber \\
&=&Z_\piro e\frac{c}{v}~\log
\left[\frac{(vt-z)+\sqrt{\frac{\robs}{\gamma^2}+(z-vt)^2}}
{(vt-z)+\sqrt{\robs+(z-vt)^2}}\right]
\eea
Using this $\Lambda({\bf r},t)$ and Equation (1b), we can calculate the 
vector potential in Coulomb gauge. In contrast to the Lorentz gauge vector 
potential, it has a component perpendicular to the $\zeze$ direction:
\bes
\bea
\left({\bf A}^\pici\right)_z~&=&~Z_\piro e~\frac{c}{v}\left[\frac{1}
{\sqrt{\robs+(vt-z)^2}}
~-~\frac{1}{\gamma \sqrt{\robs+\gamma^2(vt-z)^2}}\right]\\
\left({\bf A}^\pici\right)_\bot~&=&~Z_\piro e~\frac{c}{v}(\ro-{\bf b})
\left[\frac{1}
{\sqrt{\robs+\gamma^2(vt-z)^2}\left(\gamma(vt-z)+
\sqrt{\robs+\gamma^2(vt-z)^2}\right)}\right] \nonumber \\
&-&\left.\frac{1}{\sqrt{\robs+(vt-z)^2}\left((vt-z)+
\sqrt{\robs+(vt-z)^2}\right)}\right],\nonumber \\
\eea
while
\bea
\left({\bf A}^\piel\right)_z~&=&\frac{v}{c}\varphi^\piel(x,y,z,t)=Z_\piro e
\frac{v}{c}\frac{\gamma }{\sqrt{\robs+\gamma^2(vt-z)^2}}\\
\left({\bf A}^\piel\right)_\bot&=&0.
\eea
\ees
\smallskip
It can be verified by direct calculation that this ${\bf A}^\pici({\bf r},t)$ 
is solenoidal, $\nabla \cdot {\bf A}^\pici({\bf r},t)=0$, as specified in 
Table I for the Coulomb gauge. 

The presence of $\gamma$ in the Lorentz gauge scalar and vector potentials 
(2.3a, 2.5c) has the effect of decreasing the effective duration of the 
time-dependent impulse experienced by the target. Note that the Coulomb 
gauge potentials (2.3b, 2.5a,b) have terms that are independent of $\gamma$. 
As a result, the Coulomb gauge pulse occurs over a longer time interval than 
does the Lorentz gauge pulse, and so the Coulomb gauge pulse is more adiabatic.
This will be illustrated
in Figures 1 and 2 below.

\subsection{Finite spherically-symmetric projectile of charge $Z_{{\rm P}}e$.}

The method used in Section II.A does not work for a finite-sized projectile. 
Whereas
$\varphi^\piel({\bf r},t)$ is still given by the Equation (2.3a) (outside the
projectile), the expression for $\varphi^\pici({\bf r},t)$ is more complicated 
than Equation (2.3b). This is because the projectile, which is spherical in 
its own rest frame, appears flattened to an observer at the target, and 
Equation (2.3b) is not a solution of Poisson's equation (second row and 
second column of Table I) for a flattened charge distribution.

The conditions in the first column of Table I are Lorentz covariant. 
Thus if ($\varphi^\piel,{\bf A}^\piel$) satisfy the Lorentz gauge 
conditions in one Lorentz frame and they are subjected to a Lorentz 
transformation, then the transformed potentials will still satisfy the 
Lorentz gauge conditions. This is \textit{not} true for potentials 
satisfying the Coulomb gauge conditions given in the second column 
of Table I. If these potentials are subjected to a Lorentz transformation, 
the transformed potentials will generally \textit{not} satisfy the Coulomb 
gauge conditions. In the calculation that follows, we begin with a static 
projectile charge distribution, viewed in the projectile rest frame. In this 
case, the ordinary Coulomb potential satisfies both the Lorentz and Coulomb 
gauge conditions. However when we transform this potential to the target rest 
frame, the resulting potentials satisfy only the Lorentz conditions. If we 
want projectile potentials that satisfy the Coulomb conditions in the target 
frame, we must find a gauge transformation to take us to the Coulomb gauge 
from the potentials that have been obtained by Lorentz transformation from 
the projectile frame. 

A necessary and sufficient condition that Equations (2.1a,b) and Table I 
should be compatible is
\beq
\nabla^2\Lambda({\bf r},t)~=~\frac{1}{c}\frac{\partial}{\partial t}
\varphi^\piel({\bf r},t).
\eeq
To find a convenient expression for the right-hand side of Eq.(2.6), we 
start in the rest frame of the spherically-symmetric projectile. 
Let ${\tilde x},{\tilde y},{\tilde z}(={\tilde{\bf r}})$ be position
coordinates measured relative to the projectile center-of-mass. Then the 
scalar potential ${\tilde\varphi}({\tilde{\bf r}})$ satisfies
$$
\nabla^2{\tilde\varphi}({\tilde{\bf r}})~=~4\pi\rho_\piro({\tilde{\bf r}})
$$
whose solution can be expressed\footnote{To ensure convergence, it may be
necessary to replace $1/q^2$ by $\lim_{\eta\to 0}(1/(q^2+\eta^2))$} as
\bea
{\tilde\varphi}({\tilde{\bf r}})&=&\frac{1}{2\pi^2}\int d^3r'\int d^3q
\frac{e^{i{\bf q\cdot({\tilde{\bf r}}-{\bf r'})}}}{q^2}\rho_\piro({\bf r'})
\nonumber\\ 
&=&\frac{2}{\pi}\int d^3q\frac{e^{i{\bf q}\cdot{\tilde\bf r}}}{q^2}
\int_0^{\rm R_\piro}r'^2dr'j_0(qr')\rho_\piro(r').
\eea
It is assumed that all the projectile charge is contained within a sphere
of radius ${\rm R}_\piro$. The corresponding vector potential, 
${\tilde{\bf A}}({\tilde{\bf r}})$, is zero.

If the projectile moves so that its center is located relative to the target by
$$
{\bf r}~=~b{\hat{\bf y}}+vt\zeze~=~\ro+vt\zeze~~,
$$
an observer at the target center will measure the scalar potential
$\varphi_\piel({\bf r},t)$ to be
\beq
\varphi_\piel(x,y,z,t)~=~\gamma{\tilde\varphi}({\tilde x},{\tilde y},
{\tilde z})
\eeq
with 
\bes
\bea
{\tilde x}&=&x \label{1.5a}\\
{\tilde y}&=&y-b \label{1.5b}\\
{\tilde z}&=&\gamma(z-vt) \label{1.5c}\\
\gamma&=&(1-\frac{v^2}{c^2})^{-1/2} \label{1.5d}
\eea
\ees
Thus we can write
\bea
\varphi^\piel(x,y,z,t)&=&\frac{2\gamma}{\pi}\int d^3q
\frac{e^{i{\bf q}\cdot(x{\bf {\hat x}}+(y-b){\bf {\hat y}}+\gamma(z-vt)\zeze)}}
{q^2}
\times \int_0^{\rm R_\piro}r'^2dr'j_0(qr')\rho_\piro(r')\nonumber\\
&=&\frac{2\gamma}{\pi}\int_{-\infty}^{\infty} dq_ze^{iq_z\gamma(z-vt)}
\int d^2q_\bot
\frac{e^{i{\bf q}_\bot\cdot(\ro-{\bf b})}}{q_\bot^2+q_z^2}
\times \int_0^{\rm R_\piro}r'^2dr'j_0(qr')\rho_\piro(r')\nonumber\nonumber
\nonumber\\
&=&\frac{2}{\pi}\int_{-\infty}^{\infty} dq_ze^{iq_z(z-vt)}\int d^2q_\bot
\frac{e^{i{\bf q}_\bot\cdot(\ro-{\bf b})}}{q_\bot^2+(\frac{q_z}{\gamma})^2}
\nonumber\\
&\times& \int_0^{\rm R_\piro}r'^2dr'j_0\left(\sqrt{q^2_\bot+
(\frac{q_z}{\gamma})^2}~r'\right)\rho_\piro(r').
\eea

If equation (2.10) is used in equation (2.6), we get
$$
\nabla^2\Lambda({\bf r},t)~=~-\frac{2iv}{\pi c}\int dq_zq_z
e^{iq_z(z-vt)}\int d^2q_\bot
\frac{e^{i{\bf q}_\bot\cdot(\ro-{\bf b})}}{q_\bot^2+(\frac{q_z}{\gamma})^2}
\int_0^{\rm R_\piro}r'^2dr'j_0\left(\sqrt{q^2_\bot+
(\frac{q_z}{\gamma})^2}~r'\right)\rho_\piro(r'),
$$
whose solution can be written
$$
\Lambda({\bf r},t)~=~\frac{2iv}{\pi c}\int_{-\infty}^\infty dq_zq_z
e^{iq_z(z-vt)}\int d^2q_\bot
\frac{e^{i{\bf q}_\bot\cdot(\ro-{\bf b})}}
{q^2(q_\bot^2+(\frac{q_z}{\gamma})^2)}
\int_0^{\rm R_\piro}r'^2dr'j_0\left(\sqrt{q^2_\bot+
(\frac{q_z}{\gamma})^2}~r'\right)\rho_\piro(r')
$$
\beq
=~\frac{2i}{\pi vc}\int_{-\infty}^\infty d\omega e^{-i\omega t}\omega
e^{i\frac{\omega}{v}z}
\int d^2q_\bot
\frac{e^{i{\bf q}_\bot\cdot(\ro-{\bf b})}}
{(q_\bot^2+(\frac{\omega}{v})^2)(q_\bot^2+(\frac{\omega}{\gamma v})^2)}
\int_0^{\rm R_\piro}r'^2dr'j_0\left(\sqrt{q^2_\bot+
(\frac{\omega}{\gamma v})^2}~r'\right)\rho_\piro(r').
\eeq
in which we have replaced the integration variable $q_z$ by 
$\omega\equiv q_zv$.

We can perform the $d^2q_\bot$ integration in (2.11) by using the relations
\bes
\bea
\frac{1}{q_\bot^2+(\frac{\omega}{v})^2}\cdot\frac{1}
{q_\bot^2+(\frac{\omega}{\gamma v})^2}
&=&\frac{c^2}{\omega^2}\left[\frac{1}{q_\bot^2+(\frac{\omega}{\gamma v})^2}
-\frac{1}{q_\bot^2+(\frac{\omega}{v})^2}\right], \label{1.8a}\\
\int d^2q_\bot
\frac{e^{i{\bf q}_\bot\cdot(\ro-{\bf b})}}
{q_\bot^2+(\frac{\omega}{\gamma v})^2}~j_0\left(\sqrt{q^2_\bot+
(\frac{\omega}{\gamma v})^2}~r'\right)&=&2\pi
K_{0}\left(\frac{|\omega ||\ro-{\bf b}|}{\gamma v} \right), \label{1.8b}\\
\int d^2q_\bot
\frac{e^{i{\bf q}_\bot\cdot(\ro-{\bf b})}}
{q_\bot^2+(\frac{\omega}{v})^2}~j_0\left(\sqrt{q^2_\bot+
(\frac{\omega}{\gamma v})^2}~r'\right)&=&2\pi
K_{0}\left(\frac{|\omega ||\ro-{\bf b}|}{v} \right) 
j_0\left(i\frac{|\omega|}{c}r'\right). \label{1.8c}
\eea
\ees
Equations (2.12b,c), which are proven in the Appendix, are valid when 
$r'<|\ro-{\bf b}|$. This condition is satisfied because of our assumption 
that $b$ is large enough so that the projectile and target never overlap.
Substituting Equations (2.12a,b,c) into Equation (2.11) leads to
\bea
\Lambda({\bf r},t)&=&4i\frac{c}{v}\int_{-\infty}^\infty 
\frac{d\omega}{\omega} e^{-i\omega t}e^{i\frac{\omega}{v}z}\\
&\times&\left[\frac{Z_\piro e}{4\pi}K_{0}\left(\frac{|\omega ||\ro-{\bf b}|}
{\gamma v} \right)
~-~K_{0}\left(\frac{|\omega ||\ro-{\bf b}|}{v} \right)~\int_0^{\rm R_\piro}
r'^2dr'j_0\left(i\frac{|\omega|}{c}r'\right)
~\rho_\piro(r')\right]\nonumber 
\eea 
This $\Lambda({\bf r},t)$ will generate the transformation (2.1a,b)
between the Lorentz and the Coulomb gauges.

\section{Interaction matrix elements}

The matrix elements of the projectile-charge interaction in the two gauges
are given by
\bes
\beq
[V^\piel(t)]_{\gamma\alpha}~=~\int d^3r
\left[~[\rho_\piti({\bf r})]_{\gamma\alpha}\varphi^\piel({\bf r},t)~-~
\frac{1}{c}[{\bf J}_\piti({\bf r})]_{\gamma\alpha}\cdot{\bf A}^\piel
({\bf r},t)~\right],
\eeq
\beq
[V^\pici(t)]_{\gamma\alpha}~=~\int d^3r
\left[~[\rho_\piti({\bf r})]_{\gamma\alpha}\varphi^\pici({\bf r},t)~-~
\frac{1}{c}[{\bf J}_\piti({\bf r})]_{\gamma\alpha}\cdot{\bf A}^\pici
({\bf r},t)~\right].
\eeq
\ees
If we use Eqs. (2.1a,b) to express the differences between the potentials,
we get
\beq
[V^\pici(t)]_{\gamma\alpha}~-~[V^\piel(t)]_{\gamma\alpha}~=~
\int d^3r\left[~[\rho_\piti({\bf r})]_{\gamma\alpha}
\left(-\frac{1}{c}\frac{\partial\Lambda({\bf r},t)}{\partial t}\right)
-\frac{1}{c}[{\bf J}_\piti({\bf r})]_{\gamma\alpha}\cdot
\nabla\Lambda({\bf r},t)~\right].
\eeq
Target charge conservation (Equation (1.4)) plus Gauss' theorem, applied to 
the localized
target charge density, imply that
\bea
\int[{\bf J}_\piti({\bf r})]_{\gamma\alpha}\cdot
\nabla\Lambda({\bf r},t)d^3r&=&-\int\left(\nabla\cdot 
[{\bf J}_\piti({\bf r})]_{\gamma\alpha}\right)\Lambda({\bf r},t)d^3r
\nonumber\\
&=&i\omega_{\gamma\alpha}
\int[\rho_\piti({\bf r})]_{\gamma\alpha}\Lambda({\bf r},t)d^3r.
\eea
Thus
\beq
[V^\pici(t)]_{\gamma\alpha}~-~[V^\piel(t)]_{\gamma\alpha}~=~ -\frac{1}{c}
\int d^3r
[\rho_\piti({\bf r})]_{\gamma\alpha}\left(
\frac{\partial\Lambda({\bf r},t)}{\partial t}~+~i\omega_{\gamma\alpha}
\Lambda({\bf r},t)\right).
\eeq

\subsection{Point projectile of charge $Z_{{\rm P}}e$.}

If Equations (2.3a, 2.5c, and 1.4) are used in the expression (3.1a) for 
$[V^\piel(t)]_{\gamma\alpha}$, the result can be written in the form
\bea
[V^\piel(t)]_{\gamma\alpha}~&=&~ Z_\piro e~\int~d^3r~
{\bf J}_{\gamma\alpha}\cdot\Bigl[~\zeze\Bigl(~-\frac{(v/c^2)}
{\sqrt{(\frac{\ro-{\bf b}}{\gamma})^2+(vt-z)^2}}~+~
\frac{1}{i\omega_{\gamma\alpha}}\frac{(vt-z)}
{\bigl[(\frac{\ro-{\bf b}}{\gamma})^2+(vt-z)^2\bigr]^{3/2}}~\Bigl)
\nonumber\\
&-&\frac{1}{i\omega_{\gamma\alpha}\gamma^2}~~
\frac{\ro-{\bf b}}{\bigl[(\frac{\ro-{\bf b}}{\gamma})^2+(vt-z)^2\bigr]^{3/2}}~
\Bigl]
\eea
Similarly, $[V^\pici(t)]_{\gamma\alpha}$ can be obtained by using 
Equations (2.3b, 2.5a,b and 1.4) in (3.1b):
\bea
[V^\pici(t)]_{\gamma\alpha}~&=&~ \frac{Z_\piro e}{v}~\int~d^3r~
{\bf J_{\gamma\alpha}(r)}
\cdot
\Bigl[~\zeze~\Bigl(~\frac{1}{\gamma^2\sqrt{\frac{|\ro-{\bf b}|^2}{\gamma^2}
+(z-vt)^2}}
~-~\frac{1}{\sqrt{(\ro-{\bf b})^2+(vt-z)^2}}~\Bigr)\nonumber\\
&+&(\ro-{\bf b})\Bigl(~\frac{1}{(\ro-{\bf b})^2+(vt-z)^2+(vt-z)
\sqrt{(\ro-{\bf b})^2+(vt-z)^2}}\nonumber\\
&-&\frac{1/\gamma^2}{(\frac{\ro-{\bf b}}{\gamma})^2+(vt-z)^2+(vt-z)
\sqrt{(\frac{\ro-{\bf b}}{\gamma})^2+(vt-z)^2}}~\Bigr)~\Bigr]\nonumber\\
&+&Z_\piro e~\int~d^3r~\frac{\rho_{\gamma\alpha}}
{\sqrt{(\ro-{\bf b})^2+(vt-z)^2}}.
\eea
\medskip

\subsection{Finite spherically-symmetric projectile of charge $Z_{{\rm P}}e$.}
If we use $\Lambda({\bf r},t)$ from Equation (2.13) in Equation (3.4), we get

\bea
[V^\pici(t)]_{\gamma\alpha}~-~[V^\piel(t)]_{\gamma\alpha}
~&=&\frac{4}{v}\int_{-\infty}^\infty d\omega e^{-i\omega t}
\left(\frac{\omega_{\gamma\alpha}}{\omega}-1\right)
\int d^3r[\rho_T({\bf r})]_{\gamma \alpha} e^{i\frac{\omega}{v}z}
\\
~&\times&\left[\frac{Z_\piro e}{4\pi}
K_{0}\left(\frac{|\omega ||\ro-{\bf b}|}{\gamma v} \right)-
K_{0}\left(\frac{|\omega ||\ro-{\bf b}|}{v} \right) \int_0^{\rm R_\piro}r'^2dr'j_0\left(i\frac{|\omega|}{c}r'\right)
~\rho_\piro(r')\right]\nonumber 
\eea

The time-structure of Equation (3.7) suggests that we formulate the
expression in terms of Fourier transforms
\bes
\beq
V^{\pici,\piel}(\omega)~\equiv~\int_{-\infty}^\infty\frac{dt}{\hbar} 
e^{i\omega t}V^{\pici,\piel}(t)
\eeq
\beq
V^{\pici,\piel}(t)\equiv~\frac{\hbar}{2\pi}\int_{-\infty}^\infty d\omega
e^{-i\omega t}V^{\pici,\piel}(\omega)
\eeq
\ees
Then Equation (3.7) takes the form
$$
[V^\pici(\omega)]_{\gamma\alpha}~-~[V^\piel(\omega)]_{\gamma\alpha}~=~
\frac{8\pi}{\hbar v}\left(\frac{\omega_{\gamma\alpha}}{\omega}-1\right)
\int d^3r[\rho_T({\bf r})]_{\gamma \alpha} e^{i\frac{\omega}{v}z}
$$
\beq
\times\left[\frac{Z_\piro e}{4\pi}
K_{0}\left(\frac{|\omega ||\ro-{\bf b}|}{\gamma v} \right)-
K_{0}\left(\frac{|\omega ||\ro-{\bf b}|}{v} \right) 
\int_0^{\rm R_\piro}r'^2dr'j_0\left(i\frac{|\omega|}{c}r'\right)
\rho_\piro(r')\right].
\eeq
We see that on-shell (i.e. $\omega=\omega_{\gamma\alpha}$) interaction matrix 
elements are the same in the Coulomb and Lorentz gauges, which confirms the 
general result obtained in Section I concerning the gauge invariance of 
onshell interaction matrix elements. 

To find expressions for $[V^{\pici}(\omega)]_{\gamma\alpha}$ and
$[V^{\piel}(\omega)]_{\gamma\alpha}$ separately, we substitute 
Equations (2.5c,2.10a) into
Equation (3.1a), and use Equations (2.12b) and (3.8). 
A straightforward calculation gives
\beq
[V^\piel(\omega)]_{\gamma\alpha}~=~\frac{2Z_\piro e}{\hbar v}\int d^3r\left[
\rho_\piti({\bf r})-\frac{v}{c^2}[{\bf J}_\piti({\bf r})]_z\right]
_{\gamma\alpha}\times e^{i\frac{\omega}{v}z}
K_{0}\left(\frac{|\omega ||\ro-{\bf b}|}{\gamma v} \right)~,
\eeq
and then Equation (3.9) gives
\bea
[V^{\pici}(\omega)]_{\gamma\alpha}&=&-\frac{2Z_\piro e}{\hbar c^2}\int d^3r
\left[~[{\bf J}_\piti]_z~-~\frac{c^2}{v}\frac{\omega_{\gamma\alpha}}{\omega}
\rho_\piti~\right]_{\gamma\alpha}e^{i\frac{\omega}{v}z}
K_{0}\left(\frac{|\omega ||\ro-{\bf b}|}{\gamma v} \right)\nonumber\\
&-&\frac{8\pi}{\hbar v}\left(\frac{\omega_{\gamma\alpha}}{\omega}-1\right)
\int d^3r[\rho_\piti({\bf r})]_{\gamma\alpha}
e^{i\frac{\omega}{v}z}
K_{0}\left(\frac{|\omega ||\ro-{\bf b}|}{v} \right)\nonumber\\
&\times&\int_0^{\rm R_\piro}r'^2dr'j_0\left(i\frac{|\omega|}{c}r'\right)
\rho_\piro(r').
\eea

Finally we can proceed, as in Equation (3.6), to express the matrix elements
in terms of the current density only:
\bes
\beq
[V^\piel(\omega)]_{\gamma\alpha}~=-\frac{2Z_\piro e}{\hbar c^2}\int d^3r
[{\bf J}_\piti({\bf r})]_{\gamma\alpha}\cdot\left(\zeze+\frac{ic^2}
{v\omega_{\gamma\alpha}}\nabla\right)e^{i\frac{\omega}{v}z}
K_{0}\left(\frac{|\omega ||\ro-{\bf b}|}{\gamma v} \right),
\eeq
\bea
[V^{\pici}(\omega)]_{\gamma\alpha}&=&-\frac{2Z_\piro e}{\hbar c^2}\int d^3r
[{\bf J}_\piti({\bf r})]_{\gamma\alpha}\cdot\left(\zeze+\frac{ic^2}
{v\omega}\nabla\right)e^{i\frac{\omega}{v}z}
K_{0}\left(\frac{|\omega ||\ro-{\bf b}|}{\gamma v}\right)\nonumber\\
&&-\frac{8\pi i}{v\hbar\omega_{\gamma\alpha}}
\left(1-\frac{\omega_{\gamma\alpha}}{\omega}\right)
\int_0^{\rm R_\piro}r'^2dr'j_0\left(i\frac{|\omega|}{c}r'\right)
\rho_\piro(r')\nonumber\\
&&\times\int d^3r[{\bf J}_\piti({\bf r})]_{\gamma\alpha}\cdot\nabla
e^{i\frac{\omega}{v}z}
K_{0}\left(\frac{|\omega ||\ro-{\bf b}|}{v} \right).
\label{3.12b}
\eea
\ees
In the following sections, we will investigate significant differences
between expressions (3.12a) and (3.12b).

\section{Comparison of the structures of $V^\piel(\omega)$ and 
$V^\pici(\omega)$}
\subsection{High bombarding energy limits} 
At high bombarding energy, where $v \sim c$, the main bombarding energy
dependence enters $V^\piel(\omega)$ and $V^\pici(\omega)$ via the
$\gamma$-dependencies exhibited by Equations (3.12a) and (3.12b). Since
$$
 \lim_ {\substack{\gamma\rightarrow \infty\\v \rightarrow c}}
K_0\left(\frac{|\omega||\ro-{\bf b}|}{\gamma
v}\right)=-\log\left(\frac{|\omega||\ro-{\bf b}|}{\gamma c}\right)
~\stackrel  {\gamma \rightarrow \infty}{\longrightarrow} \log \gamma,
$$
we have
\bea
\lim_{\substack{\gamma\rightarrow \infty\\v \rightarrow c}}
[V^\piel(\omega)]_{\gamma \alpha}&=&-\frac{2 Z_{{\rm
P}}e}{\hbar c^2}~\log \gamma ~ 
\int d^3 r [{\bf J}_{{\rm T}}({\bf r})]_{\gamma \alpha}\cdot[{\bf {\hat
z}}+\frac{ic^2}{v \omega_{\gamma
\alpha}}\nabla]e^{i\frac{\omega}{v}z}\nonumber \\
&=&-\frac{2 Z_{{\rm P}}e}{\hbar c^2}~\log \gamma
~\left(1-\frac{\omega}{\omega_{\gamma \alpha}}\frac{c^2}{v^2}\right) 
\int d^3 r[{\bf J}_{{\rm T}}({\bf r})]_{\gamma \alpha}\cdot{\bf {\hat
z}} e^{i\frac{\omega}{v}z}\nonumber\\
&\stackrel{v \rightarrow c}{\longrightarrow}&-\frac{2 Z_{{\rm
P}}e}{\hbar c^2}~\log \gamma ~\left(1-\frac{\omega}{\omega_{\gamma
\alpha}}\right) 
\int d^3 r[{\bf J}_{{\rm T}}({\bf r})]_{\gamma \alpha}\cdot{\bf {\hat
z}} e^{i\frac{\omega}{v}z}.
\label{4.8}
\eea
Since the $\phi$ dependence of $[{\bf J}_{{\rm T}}({\bf r})]_{\gamma
\alpha}\cdot {\bf {\hat z}}$ is given by $e^{i(M_\gamma-M_\alpha)\phi}$,
the axial symmetry of $e^{i\frac{\omega}{v}z}$ implies that the integral
in (\ref{4.8}) vanishes unless $M_\gamma-M_\alpha~(\equiv \mu)=0$. Thus
we can have a $\log \gamma$ divergence of 
$[V^\piel(\omega)]_{\gamma \alpha}$ if $\omega \neq \omega_{\gamma \alpha}$
and $M_\gamma=M_\alpha$. The effect of this divergence on high bombarding
energy cross-sections was noted in Reference \cite{BZ}, and is illustrated in
Section VI below. 

The high-bombarding-energy behavior of $[V^\pici(\omega)]_{\gamma \alpha}$ 
is quite different. The last two lines
of Equation (3.12b) are independent of $\gamma$, and so obviously
do not diverge as $\gamma \rightarrow \infty$. The high-$\gamma$
behavior of the first line is dominated by 
$$
\lim_{\substack{\gamma\rightarrow \infty\\v \rightarrow c}}-\frac{2
Z_{{\rm P}}e}{\hbar  c^2}~\log \gamma ~ 
\int d^3 r [{\bf J}_{{\rm T}}({\bf r})]_{\gamma \alpha}\cdot[{\bf {\hat
z}}+\frac{ic^2}{v \omega}\nabla]e^{i\frac{\omega}{v}z}
$$
\begin{eqnarray*}
&=&-\frac{2 Z_{{\rm P}}e}{\hbar c^2}~\log \gamma
~\left(1-\frac{c^2}{v^2}\right) 
\int d^3 r[{\bf J}_{{\rm T}}({\bf r})]_{\gamma \alpha}\cdot{\bf {\hat
z}} e^{i\frac{\omega}{v}z}\\
&=&\frac{2 Z_{{\rm P}}e}{\hbar v^2}~\frac{\log \gamma}{\gamma^2} 
\int d^3 r[{\bf J}_{{\rm T}}({\bf r})]_{\gamma \alpha}\cdot{\bf {\hat
z}} e^{i\frac{\omega}{v}z}.
\end{eqnarray*}
The factor of $1/\gamma^2$ overpowers the $\log \gamma$
divergence, and the matrix element $[V^\pici(\omega)]_{\gamma \alpha}$ is 
seen to be finite at arbitrarily high bombarding energy.

\subsection{Multipole expansions of $V^\piel(\omega)
$ and $V^\pici(\omega) $}

It is illuminating to express Equations(3.12a) and (3.12b) in
terms of the multipole expansion given by WA:
\bes
\beq
e^{i\frac{\omega}{v}z}K_0\left(\frac{|\omega||\ro-{\bf b}|}{\gamma
v}\right)=\sum_\mu e^{-i\mu\phi_{{\bf {\rm
b}}}}K_\mu\left(\frac{|\omega|b}{\gamma v}\right)\sum_\lambda {\cal 
G}_{\lambda \mu}~
j_\lambda\left(\frac{|\omega|}{c}r\right)Y^\lambda_\mu({\bf {\hat r}})\\
\label{4.9a}
\eeq
with ${\cal G}_{\lambda \mu}$ defined by
\bea 
{\cal G}_{\lambda\mu}&\equiv&
\frac{i^{\lambda+\mu}}{(2\gamma)^\mu}\left(\frac{\omega}
{|\omega|}\right)^{\lambda-\mu}\left(\frac{c}{v}\right)
^\lambda\sqrt{4\pi(2\lambda+1)(\lambda-\mu)!(\lambda+\mu)!}
\nonumber\\
&\times& \sum_n\frac{1}{(2\gamma)^{2n}(n+\mu)!n!(\lambda-\mu-2n)!}
\label{4.9b}.
\eea
\ees 
To expand the third line of Equation (3.12b), we also need the
$\gamma \rightarrow 1$ limits of Equations (\ref{4.9a}, \ref{4.9b}). To
perform these
limits, without affecting the value of $v$, we allow $c \rightarrow
\infty$, and obtain
\bea
e^{i \frac{\omega}{v}z}K_0\left(\frac{|\omega||\ro-{\bf
b}|}{v}\right)&=&\sum_\mu e^{-i\mu \phi_{{\bf
b}}}K_\mu\left(\frac{|\omega| b}{v}\right)\sum_\lambda\sqrt{\frac{4
\pi}{2 \lambda
+1}}\frac{i^{\lambda+\mu}}{\sqrt{(\lambda+\mu)!(\lambda-\mu)!}}\nonumber
\\
&\times&
\left(\frac{|\omega|}{\omega}\right)^{\lambda-\mu}\left(\frac{\omega 
r}{v}\right)^\lambda Y^\lambda_\mu({\bf {\hat  r}}).
\label{4.10}
\eea
If we apply Equations (\ref{4.9a}) and (\ref{4.10}) to Equation
(3.12b), we find 
\begin{equation}
[V^\pici(\omega)]_{\gamma \alpha} =\frac{2 Z_{{\rm
P}}e}{\hbar
v}\sum_\mu e^{-i\mu\phi_b}
\cdot\sum_{\lambda=|\mu|}^\infty\left[~\left(X^\lambda_\mu(E)+
X^\lambda_\mu(M)\right)K_\mu\left(\frac{|\omega|b}{\gamma
v}\right) +X^\lambda_\mu(S)K_\mu\left(\frac{|\omega|b}{v}\right) ~\right],
\label{4.11}
\end{equation}
where
\bes   
\bea
X^\lambda_\mu(E)&\equiv& \frac{iv}{c\hbar\omega}\left[\frac{{\cal
G}_{\lambda-1,\mu}}{\lambda}\sqrt{\frac{\lambda^2-\mu^2}{(2
\lambda+1)(2\lambda-1)}}+\frac{{\cal
 G}_{\lambda+1,\mu}}{\lambda+1}\sqrt{\frac{(\lambda+1)^2-\mu^2}{(2
\lambda+1)(2\lambda+3)}}\right]\nonumber\\
& \times & \int d^3 r [{\bf J}_{{\rm T}}]_{\gamma \alpha}({\bf r})\cdot
(\nabla\times
 {\bf L})j_\lambda(\frac{\omega}{c}r)Y^\lambda_\mu({\bf {\hat r}})
\label{4.12a}\\
X^\lambda_\mu(M)&\equiv& -\frac{v \mu}{c^2\hbar}\frac{{\cal
G}_{\lambda,\mu}}{\lambda(\lambda+1)}
\times\int d^3 r [{\bf J}_{{\rm T}}]_{\gamma \alpha}({\bf r})\cdot
{\bf L}j_\lambda(\frac{\omega}{c}r)Y^\lambda_\mu({\bf {\hat r}})
\label{4.12b} \\
X^\lambda_\mu(S)&\equiv&\left(1-\frac{\omega_{\gamma
\alpha}}{\omega}\right)\frac{4 \pi}{Z_{{\rm P}}e} \int_0^\infty {\tilde
r}^2 d{\tilde r} {\tilde \rho}_{{\rm P}}({\tilde r})j_0\left(i{\tilde
r}\frac{\omega}{c}\right) \sqrt{\frac{4 \pi}{2
\lambda+1}}\frac{i^{\lambda+\mu}}{\sqrt{(\lambda+\mu)!(\lambda-\mu)!}}
\nonumber \\
&\times& \left(\frac{|\omega|}{\omega}\right)^{\lambda-\mu}\frac{1}{i
\omega_{\gamma \alpha}}\left(\frac{\omega}{v}\right)^\lambda 
\times \int d^3r[ {\bf J}_{{\rm T}}({\bf r})]_{\gamma \alpha} \cdot
\nabla r^\lambda Y^\lambda_\mu({\bf {\hat r}}).
\label{4.12c}
\eea
\ees
On the other hand, substituting Equation (\ref{4.9a}) into Equation
(3.12a) gives
\beq
[V^\piel(\omega)]_{\gamma \alpha} =\frac{2 Z_{{\rm
P}}e}{\hbar
v}\sum_\mu e^{-i\mu\phi_{{\bf {\rm b}}}}
\cdot\sum_{\lambda=|\mu|}^\infty~\left(X^\lambda_\mu(E)+
X^\lambda_\mu(M)+X^\lambda_\mu(G)\right)
K_\mu\left(\frac{|\omega|b}{\gamma v}\right).
\label{4.13}
\eeq

$X^\lambda_\mu(E)$ and $X^\lambda_\mu(M)$ in Equation (\ref{4.13}) are
the same as in Equations (\ref{4.12a} and \ref{4.12b}), but
$X^\lambda_\mu(G)$ is defined by
\beq
X^\lambda_\mu(G)\equiv\left(1-\frac{\omega_{\gamma
\alpha}}{\omega}\right)\frac{{\cal G}_{\lambda \mu}}{i \omega_{\gamma
\alpha}}\int d^3 r [{\bf J}_{{\rm T}}({\bf r})]_{\gamma \alpha} \cdot
\nabla j_\lambda\left(\frac{|\omega|}{c}r\right)Y^\lambda_\mu({\bf {\hat
r}}).
\label{4.14}
\eeq

\subsection{Monopole matrix elements}

Inspection of Equation (\ref{4.11}) and (\ref{4.12a}, \ref{4.12b} and
{\ref{4.12c}) shows that $[V^\pici(\omega)]_{\gamma
\alpha}$ can be expressed as
\beq
[V^\pici(\omega)]_{\gamma \alpha}=\int d^3r[{\bf
J}_{{\rm T}}({\bf r})]_{\gamma \alpha}\cdot {\bf Q}^\pici({\bf r})
\label{4.15}
\eeq
where ${\bf Q}^\pici({\bf r})$ is a linear combination of $
(\nabla\times
 {\bf L})j_\lambda(\frac{\omega}{c}r)Y^\lambda_\mu({\bf {\hat r}}),~ 
{\bf L}j_\lambda(\frac{\omega}{c}r)Y^\lambda_\mu({\bf {\hat r}})$, and
$\nabla r^\lambda Y^\lambda_\mu({\bf {\hat r}})$. Since 
$$
\nabla\cdot(\nabla\times
 {\bf L})j_\lambda(\frac{\omega}{c}r)Y^\lambda_\mu({\bf {\hat
r}})=0,~~~\nabla \cdot{\bf
L}j_\lambda(\frac{\omega}{c}r)Y^\lambda_\mu({\bf {\hat r}})=0,~~~\nabla
\cdot \nabla r^\lambda Y^\lambda_\mu({\bf {\hat r}})=0,
$$
it follows that 
\beq
 \nabla \cdot {\bf Q}^\pici({\bf r})=0.
\label{4.16}
\eeq
Now suppose that $\phi_\alpha$ and $\phi_\gamma$ are both states of total 
angular momentum zero ($J=0$). Then 
$[{\bf J}_{{\rm T}}({\bf r})]_{\gamma \alpha}$ is a spherically-symmetric 
vector field (a \textit{central} field), and can be written as the gradient 
of a spherically-symmetric scalar field:
$$
[{\bf J}_{{\rm T}}({\bf r})]_{\gamma \alpha}=\nabla \psi(r).
$$
Equations (\ref{4.15}) and (\ref{4.16}) imply that
$$
[V^\pici(\omega)]_{\gamma \alpha}=\int d^3 r \nabla
\psi(r)\cdot {\bf Q}^\pici({\bf r}) = -\int d^3 r \psi(r) \nabla
\cdot{\bf Q}^\pici({\bf r}) = 0.
$$
so that the matrix elements of $V^\pici$ between any two $J=0$
states vanish identically. Thus an excited $J=0$ state, in a nucleus
with a
$J=0$ ground state, can only  be populated indirectly, via a multi-step
process. However, if $[V^\piel(\omega)]_{\gamma
\alpha}$ is expressed as
\beq
[V^\piel(\omega)]_{\gamma \alpha}=\int d^3r[{\bf
J}_{{\rm T}}({\bf r})]_{\gamma \alpha}\cdot {\bf Q}^\piel({\bf r}),
\label{4.17}
\eeq
then one of the components in the expansion of ${\bf Q}^\piel({\bf
r})$ is 
$\nabla j_\lambda(|\omega|r/c)Y^\lambda_\mu({\bf
{\hat r}})$ (see Equations (\ref{4.13},\ref{4.14})). But 
$$
\nabla\cdot\nabla
j_\lambda\left(\frac{|\omega|}{c}r\right)Y^\lambda_\mu({\bf {\hat
r}})=\nabla^2
j_\lambda\left(\frac{|\omega|}{c}r\right)Y^\lambda_\mu({\bf {\hat
r}})=-\left(\frac{\omega}{c}\right)^2~j_\lambda\left(\frac{|\omega|}{c}r
\right)Y^\lambda_\mu({\bf
{\hat
r}}),
$$
which is not identically zero. Thus $\nabla \cdot{\bf Q}^\piel({\bf
r})$ is not identically zero and there is no reason to expect matrix
elements of $V^\piel$ between $J=0$ states to vanish. A calculation
based on $V^\piel$ implies the possibility of one-step population
of an excited $J=0$ state from a $J=0$ ground state.

\section{Numerical comparisons of matrix elements of $V^\piel$ and $V^\pici$}

The following numerical comparisons refer to Coulomb excitation of the
giant dipole resonance (GDR) in a  $^{40}$Ca target by a $^{208}$Pb
projectile. The target transition current density matrix elements
$[{{\bf J}_{{\rm T}}({\bf r})}]_{\gamma \alpha}$ needed in Equations
(4.5) and (4.7) are calculated using Brink's model \cite{Bri}
of the
GDR, in which unexcited proton and neutron spheres undergo harmonic
oscillations relative to each other. The radial densities of the spheres
are obtained by filling the lowest available shell-model single-particle
states. The relative harmonic oscillations of the proton and neutrons
spheres are characterized by the eigenstates $\phi^{n,\ell}_m({\bf r}_{{\rm
pn}})$ of a three-dimensional harmonic oscillator. The ground state is
$\phi^{0,0}_0$. The degenerate first excited states that can be
populated by Coulomb excitation are $\phi^{0,1}_0$ and
$\left(\phi^{0,1}_1+\phi^{0,1}_{-1}\right)/\sqrt{2}$, the
combination of $\phi^{0,1}_{\pm 1}$ symmetric under reflection across
the reaction plane. Further details on the calculation of matrix elements using the Brink model eigenstates are given in reference \cite{BZ}.
 
In the calculations represented in Figure 1, it has been assumed that
the 82 protons of $^{208}$Pb are distributed uniformly within a sphere
of radius 7.5 fm (as seen by an observer moving with the projectile).
The parameters used for the calculations imply that $b=20$ fm and
$\omega_{\gamma \alpha}=11.7$ MeV$/\hbar$. It can be verified by
inspection of Figure 1 that, in every case, $V^\pici$
and $V^\piel$ agree at $\omega=11.7$ MeV$/\hbar$, as
required by Equation (3.9).

In Figure 1, $\mu=0$ refers to the transition
$\phi^{0,0}_0\rightarrow \phi^{0,1}_0$ and $\mu=1$ refers to the
transition $\phi^{0,0}_0\rightarrow
\left(\phi^{0,1}_1+\phi^{0,1}_{-1}\right)/\sqrt{2}$. In every
case, the solid line gives $V^\pici(\omega)$, whereas
the dashed line gives $V^\piel(\omega)$.

The first observation concerning Figure 1 is that at a bombarding energy
of E/A=100 MeV, there is little difference between $V^\pici$ and
$V^\piel$. However, at E/A=10 GeV, the difference is pronounced.
The most striking difference is the very strong increase with bombarding
energy of $V^\piel$ for the $\mu=0$ transition. This is an
expression of the logarithmic divergence (with increasing $\gamma$) of
the $\mu=0$ matrix elements of $V^\piel$, as shown in Equation
(\ref{4.8}). It is clear that, at high bombarding energy, calculations
using $V^\piel$ will ascribe a much higher role to $\mu=0$
transitions than will calculations using $V^\pici$.

Figure 1 also shows that, at high bombarding energy, the
use of the Coulomb gauge leads to an interaction that is more adiabatic
than predicted by the Lorentz gauge. The wider spread
of $V^\piel(\omega)$, as a function of $\omega$,
compared to the spread of $V^\pici(\omega)$, shows
that the impulse in the Lorentz gauge is 
sharper than the Coulomb gauge impulse. For example, Figure 2 shows the same 
comparison as in the $E/A=10$ GeV, $\mu=1$ plot of Figure 1, but in the time 
domain. The sharpness of the pulse provided by $V^\piel(t)$, compared with 
that provided by $V^\pici(t)$, is evident. Note that in this particular case, 
$[V^\piel(-\omega)]_{\gamma \alpha}=[V^\piel(\omega)]_{\gamma \alpha}$ which 
has the consequence that $[V^\piel(t)]_{\gamma \alpha}$ is real. However, 
this is not true of $[V^\pici(\omega)]_{\gamma \alpha}$, and thus 
$[V^\pici(t)]_{\gamma \alpha}$ has both real and imaginary parts.  
 
\section{Gauge-dependent effects in coupled-channel calculations.}

The most significant test of the differences between $V^\piel$ and
$V^\pici$ is in the calculation of RCE cross-sections, since the
cross-section is the point where theory and experiment intersect. In
this Section, we compare cross-sections calculated with these two
interactions, for the $^{208}$Pb + $^{40}$Ca system described in Section V.
We have performed coupled-channel integrations of the time-dependent 
Schr\"odinger 
equation. The target states included in the calculation
span the 0-phonon, 1-phonon, and 2-phonon states of the $^{40}$Ca GDR.
The methods used to do the numerical Fourier transform needed in
Equation (3.8b), and to integrate the coupled equations, are
described in Reference \cite{BZ}. We also describe there the integrations over
impact parameter needed to calculate the cross-section.

Figure 3 compares the calculated cross-sections for the population of
the six reflection-symmetric 1- and 2-phonon states that can be reached
via RCE, as functions of the kinetic energy of the $^{208}$Pb projectile
nucleus. In every case, a solid line is used to show the result of the
calculation using $V^\pici(t)$, and a dashed line is used to show
the result of the corresponding calculation using $V^\piel(t)$.

For the one-phonon states, the situation is similar to that shown in Figure 1. 
The two sets of calculations agree at low bombarding energy 
($E/A \stackrel{\sim}{<} 1$ GeV). However, at high bombarding energy 
($E/A \stackrel{\sim}{>} 5$ GeV), $V^\piel(t)$
predicts much greater cross-sections for the population of the $J=1,M=0$ state
than does $V^\pici(t)$. The situation is more complicated for the two-phonon 
states. The cross-sections are much smaller than for the one-phonon states, 
and depend upon multiple excitation processes. Also, the truncation of our 
calculation at two phonons introduces an additional element of uncertainty 
into our two-phonon cross-sections. However, it is noteworthy that 
the 2-phonon $|M|=1$ state is also strongly favored by $V^\piel(t)$ at high 
bombarding
energy, compared to $V^\pici(t)$.  The reason is that this state is mostly 
populated in two-step processes, such as  
$$
(J=M=0) \rightarrow (J=1,M=0) \rightarrow (J=2,|M|=1)
$$
and
$$
(J=M=0) \rightarrow (J=1,|M|=1) \rightarrow (J=2,|M|=1).
$$ 
In both cases, a $\Delta M=0$ transition is involved, and it is the
logarithmic divergence (with increasing $\gamma$) of the $\Delta M=0$
transition amplitude that leads to the strongly increasing cross-section
when $V^\piel(t)$ is used. But the $J=2,|M|=2$ state is reached
mainly by 
$$
(J=M=0) \rightarrow (J=1,|M|=1) \rightarrow (J=2,|M|=2),
$$ 
in which there is no $\Delta M=0$ step, and so the $J=2,|M|=2$ state is
not strongly favored by $V^\piel(t)$ at high bombarding energy. 
Thus we see that for $E/A \stackrel{\sim}{>} 5$ GeV, $V^\pici(t)$
and $V^\piel(t)$ predict very different cross-sections when used in
truncated coupled-channel analyses. A calculation using $V^\piel(t)$ will
predict stongly enhanced cross-sections for populating any state that
can be reached via a one-step or two-step process involving a $\Delta
M=0$ transition.

The stongest transition illustrated in Figure 3 populates the one-phonon 
$|M|=1$ state. It is predominantly a one-step transition, and so is well 
described by the Born approximation. According to Equation (\ref{4.11}) 
or (\ref{4.13}), this implies that the $b$-dependence of the transition 
probability is given by  
$$
\left[~K_1\left(\frac{\omega_{{\rm on-shell}}  b}{\gamma
v}\right)~\right]^2
$$
whose the integral over impact parameter diverges in the 
$\gamma\rightarrow \infty$ limit (cf Equations (3.1), (3.6) of Reference 
\cite{WA}.) The physical reason for this divergence is that as $\gamma$
increases and the electromagnetic pulse becomes more strongly retarded,
it becomes flatter and its influence extends for a longer distance away
from the trajectory of the projectile. Thus, in the integral over $b$,
larger values of $b$ play a more important role as $\gamma$ increases,
and in the $\gamma \rightarrow \infty$ limit, the $b$-integral diverges.
This occurs only for $\Delta M=\pm 1$, because the flat pulse is
spatially axially symmetric ($\Delta M=0$) and the intrinsic spin of the
photon transfers $\Delta M=\pm 1$. The one-phonon $|M|=1$ cross-section shown 
in Figure 3 exhibits this logarithmic-like increase with bombarding energy.

However, the $\log \gamma$ divergence of the off-shell $\mu=\Delta M=0$ 
matrix element, discussed in Section IV.C,
is a much more serious divergence. It occurs for
\textit{every} $b$, and thus the $b$-integrated cross-section would be a
divergence of still higher order. This is the divergence that is
introduced when the Lorentz gauge is used in a
coupled-channel calculation, a divergence which is not present when the
Coulomb gauge is used.

The expression for $V^\piel(\omega)$ given by Equation
(3.12a) depends upon the total projectile charge, but not on the
radial dependence of this charge. However there \textit{is} a dependence
on the projectile radial charge density in $V^\pici(\omega)$ given by Equation 
(3.12b). Fortunately, this
dependence is very weak. The calculations using $V^\pici(\omega)$ whose 
results are shown in Figure 3, were done using a
radial charge distribution which was constant from the center out to 7.5
fm. By way of comparison, we have repeated these calculations, assuming
that all the charge of the $^{208}$Pb nucleus is concentrated at its
center. In all cases, we found that the calculated cross-sections
changed by no more than a few tenths of a percent. As might be expected,
the transition  amplitudes are more sensitive to projectile radial
density at small impact parameters. However, when the integration over
all impact parameters is done, the residual effect on the cross-section
of changes in the projectile radial charge density is very small. 
\section{Conclusions and discussion}
We have studied the relationship between the electromagnetic potentials in 
Lorentz gauge and Coulomb gauge for the fields encountered in relativistic 
Coulomb excitation, and have found expressions in the two gauges for the 
interaction between projectile and target. At bombarding energies above 
about 2 GeV per nucleon, we have found significant differences in excitation 
cross-sections when the two gauges are used in coupled-channel time-dependent 
calculations, especially for processes involving $\Delta 
M=0$ transitions. Since there is no \textit{a priori} reason for using one 
gauge rather than another, this lack of gauge invariance reveals a weakness of 
the coupled-channel time-dependent approach to relativistic Coulomb excitation.
It demonstrates that 
the common practice of relying totally on calculations in the Lorentz gauge 
cannot be justified.

We have discussed one cause of lack of gauge invariance, the truncation of 
the set of target states used in the coupled-channel calculation. Another 
cause is revealed by consideration of the Hamiltonian for a target charged 
particle moving in the fields $
\left(\varphi({\bf r},t),~{\bf A(r},t)\right)$ ~produced by the projectile:
\begin{eqnarray}
H&=&\frac{1}{2m}\left({\bf p}-\frac{e{\bf A(r},t)} {c}\right)^2+ 
e\varphi({\bf r},t) ~+~U_{{\rm nuc}}({\bf r}) \nonumber \\
~&=&\frac{1}{2m}{\bf p}^2-\frac{e}{2mc}\left(~{\bf p \cdot A(r},t)+
{\bf A(r},t){\bf \cdot p}~\right)+e\varphi({\bf r},t) \nonumber \\ 
&+&\frac{e^2}{2 m c^2}{\bf A(r},t){\bf \cdot A(r},t)~+~U_{{\rm nuc}}({\bf r})
\end{eqnarray}
The terms in Equation (7.1) that are linear in $e$,
$$
-\frac{e}{2mc}\left(~{\bf p \cdot A(r},t)+{\bf A(r},t){\bf \cdot p}~\right)
+e\varphi({\bf r},t), 
$$ 
give rise to the interaction (1.3) that has been the basis of this study. 
However the term quadratic in $e$,
$$
\frac{e^2}{2 m c^2}{\bf A(r},t){\bf \cdot A(r},t),
$$
has not been included in our calculation, or in other studies published to 
date on relativistic Coulomb excitation. But the equation
$$
\left({\bf p}-\frac{e({\bf A(r},t)+\nabla \Lambda({\bf r},t))} {c}
\right)^2~e^{i\frac{e \Lambda({\bf r},t)}{\hbar c}}\psi({\bf r},t)=
e^{i\frac{e \Lambda({\bf r},t)}{\hbar c}}\left({\bf p}-\frac{e{\bf A(r},t)} 
{c}\right)^2~\psi({\bf r},t),
$$
which is an essential component in the argument about the gauge invariance of 
the observable consequences of the Schr\"odinger equation, relies for its 
validity on the presence of the ${\bf A \cdot A}$ term. Thus it is clear 
that no gauge invariant theory
 of relativistic Coulomb excitation can be constructed without inclusion of 
this term. We will address this issue in a forthcoming publication.
  
\appendix*
\section{Two integral relations}
Define $I({\tilde r})$ by
\beq
I({\tilde r})\equiv\int\frac{d^2{\bf q}_\bot}{q_\bot^2+\lambda_1^2}~e^{i
{\bf q}_\bot\cdot (\ro-{\bf b}) }~j_0\left({\tilde
r}\sqrt{q_\bot^2+\lambda_2^2}\right).
\label{B.1}
\eeq
Operate on this expression with $\nabla^2_{\bf {\tilde r}}$:
\bea
\nabla^2_{\bf {\tilde r}} I({\tilde r})&=&\int\frac{d^2{\bf
q}_\bot}{q_\bot^2+\lambda_1^2}~e^{i {\bf q}_\bot\cdot (\ro-{\bf
b})}~\nabla^2_{\bf {\tilde r}} j_0\left({\tilde
r}\sqrt{q_\bot^2+\lambda_2^2}\right)\nonumber \\
&=&-\int d^2{\bf
q}_\bot\frac{q_\bot^2+\lambda_2^2}{q_\bot^2+\lambda_1^2}~ e^{i {\bf
q}_\bot\cdot (\ro -{\bf b})}~j_0\left({\tilde
r}\sqrt{q_\bot^2+\lambda_2^2}\right)\nonumber \\
&=& -\int d^2{\bf q}_\bot \left(1+
\frac{\lambda_2^2-\lambda_1^2}{q_\bot^2+\lambda_1^2}\right) ~e^{i {\bf
q}_\bot\cdot (\ro-{\bf b})}~j_0\left({\tilde
r}\sqrt{q_\bot^2+\lambda_2^2}\right)\nonumber \\
&=&-\int d^2{\bf q}_\bot e^{i {\bf q}_\bot\cdot (\ro-{\bf b})
}~j_0\left({\tilde r}\sqrt{q_\bot^2+\lambda_2^2}\right)\nonumber \\
&-&(\lambda_2^2-\lambda_1^2)\int\frac{d^2{\bf
q}_\bot}{q_\bot^2+\lambda_1^2}~e^{i {\bf q}_\bot\cdot (\ro-{\bf
b})}~j_0\left({\tilde r}\sqrt{q_\bot^2+\lambda_2^2}\right). 
\label{B.2}
\eea
In the first term of Equation (\ref{B.2}), replace $j_0\left({\tilde
r}\sqrt{q_\bot^2+\lambda_2^2}\right)$ by
$$
j_0\left({\tilde r}\sqrt{q_\bot^2+\lambda_2^2}\right)=\frac{1}{4
\pi}\int \sin {\tilde \theta} d {\tilde \theta} d{\tilde \phi}e^{-i
({\bf q}_\bot \cdot {\tilde \ro}+\lambda_2 {\tilde z})}
$$
where ${\tilde \ro}\equiv {\tilde r}\sin {\tilde \theta}(\cos {\tilde
\phi}{\bf {\hat x}}+\sin {\tilde \phi}{\bf {\hat y}})$ and ${\tilde
z}={\tilde r}\cos {\tilde \theta}$. Then Equation (\ref{B.2}) becomes
\bea
\nabla^2_{\bf {\tilde r}} I({\tilde r})&=&-\frac{1}{4 \pi}\int \sin
{\tilde \theta} d {\tilde \theta} d {\tilde \phi} \int d^2 {\bf
q}_\bot~e^{i{\bf q}_\bot\cdot(\ro-{\bf b}-{\tilde
\ro})}~e^{-i\lambda_2{\tilde z}}\nonumber\\
&-&(\lambda_2^2-\lambda_1^2)\int \frac{d^2 {\bf
q}_\bot}{q_\bot^2+\lambda_1^2}~e^{i{\bf q}_\bot\cdot(\ro-{\bf
b})}~j_0\left({\tilde r}\sqrt{q_\bot^2+\lambda_2^2}\right)\nonumber \\
\nabla^2_{\bf {\tilde r}} I({\tilde
r})+(\lambda_2^2-\lambda_1^2)I({\tilde r})&=&-\frac{(2 \pi)^2}{4
\pi}\int \sin {\tilde \theta} d{\tilde \theta} d{\tilde \phi}e^{-i
\lambda_2 {\tilde z}}\delta(\ro-{\bf b}-{\tilde \ro}).
\label{B.3}
\eea
In our application, $|\ro|$ is bounded by the radius of the target and
$|{\tilde \ro}|$ is bounded by the radius of the projectile. The
condition that the projectile and target do not overlap implies that
$|\ro-{\bf b}|>|{\tilde \ro}|$ for every orientation ${\tilde
\theta},{\tilde \phi}$, and so $\ro-{\bf b}-{\tilde \ro}$ is never
zero. Thus the $\delta$-function on the right-hand side of Equation
(\ref{B.3}) vanishes, and we get
\beq
\nabla^2_{\bf {\tilde r}} I({\tilde r})+ (\lambda_2^2-\lambda_1^2)
I({\tilde r})=0.
\label{B.4}
\eeq 

We have two applications of Equation (\ref{B.4}). In one of them,
$\lambda_1^2=\lambda_2^2=(\omega/(\gamma v))^2$. In the other,
$\lambda_1^2=(\omega/v)^2,~\lambda_2^2=(\omega/(\gamma
v))^2$. Thus in both cases, 
$(\lambda_2^2-\lambda_1^2) \leq 0$, and we can write the general
solution of Equation (\ref {B.4}) as
\beq
I({\tilde r})= \alpha ~j_0 \left(i{\tilde
r}\sqrt{\lambda_1^2-\lambda_2^2}\right) ~+~\beta ~n_0\left(i{\tilde
r}\sqrt{\lambda_1^2-\lambda_2^2}\right), 
\label{B.5}
\eeq
where $\alpha$ and $\beta$ are independent of ${\tilde r}$. The first
term in Equation (\ref{B.5}) is finite at ${\tilde r}=0$, whereas the
second term diverges there. But if we set ${\tilde r}=0$ in the
definition (\ref{B.1}), we get
\beq
I(0)=\int\frac{d^2{\bf q}_\bot}{q_\bot^2+\lambda_1^2}~e^{i {\bf
q}_\bot\cdot (\ro-{\bf b}) }=2 \pi K_{0}(|\lambda_1||\ro-{\bf b}|)
\label{B.6}
\eeq
which is finite in our $|\ro-{\bf b}|\geq 0$ situation. Thus $\beta$ in
Equation ({\ref{B.5}) vanishes and
$$
I(0)=\alpha j_0(0)=\alpha=2 \pi K_{0}(|\lambda_1||\ro-{\bf b}|),
$$
leading to
\beq 
I({\tilde r})\equiv\int\frac{d^2{\bf q}_\bot}{q_\bot^2+\lambda_1^2}~e^{i
{\bf q}_\bot\cdot (\ro-{\bf b}) }~j_0\left({\tilde
r}\sqrt{q_\bot^2+\lambda_2^2}\right)=2 \pi K_{0}(|\lambda_1||\ro-{\bf
b}|)
~j_0 \left(i{\tilde r}\sqrt{\lambda_1^2-\lambda_2^2}\right). 
\eeq
\label{B.7}
Thus we get the two special cases:
\bes
\bea
\int\frac{d^2{\bf q}_\bot}{q_\bot^2+(\frac{\omega}{\gamma v})^2}~e^{i
{\bf q}_\bot\cdot (\ro-{\bf b}) }~j_0\left({\tilde
r}\sqrt{q_\bot^2+(\frac{\omega}{\gamma v})^2}\right)=2 \pi
K_{0}\left(\frac{|\omega ||\ro-{\bf b}|}{\gamma v} \right)\\
\label{B.8a}
\int\frac{d^2{\bf q}_\bot}{q_\bot^2+(\frac{\omega}{v})^2}~e^{i {\bf
q}_\bot\cdot (\ro-{\bf b}) }~j_0\left({\tilde
r}\sqrt{q_\bot^2+(\frac{\omega}{\gamma v})^2}\right)=2 \pi
K_{0}\left(\frac{|\omega||\ro-{\bf b}|}{v} \right)
~j_0 \left(i{\tilde r}\frac{|\omega|}{c}\right). 
\label{B.8b}
\eea
\ees
These equations are valid when ${\tilde r}<|\ro-{\bf b}|$.

\newpage

\centerline{Figure Captions}
\medskip

Figure 1. Plots of matrix elements of $V^{\pici}(\omega)$ (solid lines) and $V^{\piel}(\omega)$ (dashed lines) connecting 
the $^{40}$Ca ground state to the one-phonon giant dipole resonance states. 
The projectile is a $^{208}$Pb nucleus with bombarding energies per nucleon 
specified within the figure frames. $\mu=0$ corresponds to the $M=0$ one-phonon state; $\mu=1$ to the reflection-symmetric $|M|=1$ state.

Figure 2. Comparison of $V^\pici(t)$ (solid line) and $V^\piel(t)$ 
(dashed line), corresponding to the $E/A=10$ GeV, $\mu=1$ example of Figure 1.

Figure 3. Calculated coupled-channel Coulomb excitation cross-sections for 
one- and two-phonon states of the GDR in $^{40}$Ca with $^{208}$Pb projectiles.
The solid curves correspond to calculations in which the Coulomb gauge has been used, 
the dashed curves to calculations with the Lorentz gauge.

\end{document}